\documentclass[preprint2,10.5pt]{aastex}
\usepackage{epsfig}
\begin{document}
\title{\large{\rm{THE ONSET OF CHAOS IN PULSATING VARIABLE STARS}}}
\author{David G. Turner$^{1,5}$, Leonid N. Berdnikov$^{2,5}$, John R. Percy$^3$, Mohamed Abdel-Sabour Abdel-Latif$^4$}
\affil{$^1$ Saint Mary's University, Halifax, Nova Scotia, Canada}
\affil{$^2$ Sternberg Astronomical Institute, Moscow, Russian Federation}
\affil{$^3$ Erindale College, University of Toronto, Erindale, Ontario, Canada}
\affil{$^4$ National Research Institute of Astronomy and Geophysics, Helwan, Egypt}
\affil{$^5$ Visiting Astronomer, Harvard College Observatory Photographic Plate Stacks}
\email{\rm{turner@ap.smu.ca}}

\begin{abstract}
Random changes in pulsation period occur in cool pulsating Mira variables, Type A, B, and C semiregular variables, RV Tauri variables, and in most classical Cepheids. The physical processes responsible for such fluctuations are uncertain, but presumably originate in temporal modifications of the envelope convection in such stars. Such fluctuations are seemingly random over a few pulsation cycles of the stars, but are dominated by the regularity of the primary pulsation over the long term. The magnitude of stochasticity in pulsating stars appears to be linked directly to their dimensions, although not in simple fashion. It is relatively larger in M supergiants, for example, than in short-period Cepheids, but is common enough that it can be detected in visual observations for many types of pulsating stars. Although chaos was discovered in such stars 80 years ago, detection of its general presence in the group has only been possible in recent studies.
\end{abstract}


\section{Introduction}
\begin{figure*}[!t]
\centerline{
\epsfig{file=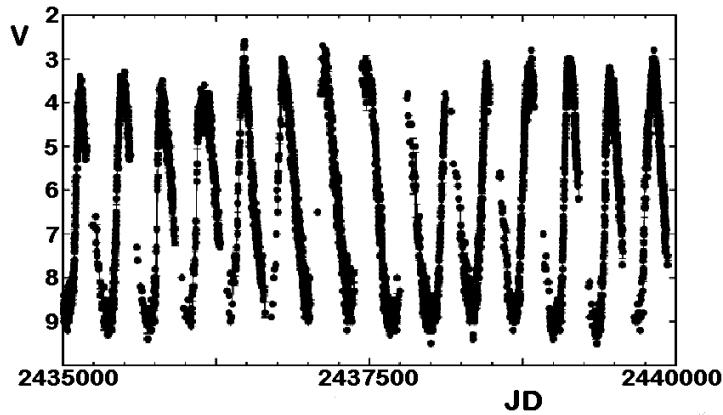, scale=0.32}
}
\caption{\small{The visual light curve of Mira between JD 2435000 and JD 2440000 from AAVSO measurements.}}
\label{fig1}
\end{figure*}

A well known problem in variable star studies is that it is impossible to predict exact moments for light maximum in some late-type pulsating variables, such as Miras and semi-regular variables, or to predict their amplitude on any given cycle (see Fig.~\ref{fig1} for AAVSO (American Association of Variable Star Observers) observations of the name star, $o$ Ceti). The cyclical light patterns displayed in such stars are reasonably well defined over long time intervals and can be approximated closely with linear ephemerides, but the regularity of their pulsation is typically marked by other effects best revealed through careful O--C analysis.

  A common complication is that of ``random'' fluctuations in pulsation period for a star from one cycle to another. Many years ago Eddington \& Plakidis (1929) developed an interesting technique for establishing the importance of random fluctuations in pulsation period for Mira variables, and it has been revived frequently in recent years (Percy \& Hale 1998; Percy \& Colivas 1999, Percy et al. 1993, 2003, 2007) in order to establish the importance of random changes in period for other Mira variables as well as for other types of cool and hot pulsating variables.

\begin{figure}[!t]
\centerline{
\epsfig{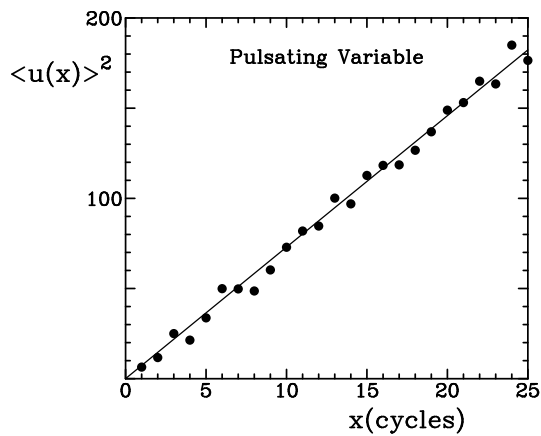}
}
\caption{\small{A schematic representation of an Eddington-Plakidis test for a pulsating variable with a randomness factor of $e=2.7$ days and uncertainties of $a=0.15$ day in measured times of light maximum.}}
\label{fig2}
\end{figure}

  The technique has been described previously in these pages (Turner \& Berdnikov 2001), and involves computing, without regard to sign, the average accumulated time delays $\langle u(x) \rangle$ between light maxima separated by {\it x} cycles. If the deviations in the observed times of light maxima from their predicted times are dominated by random fluctuations in period, then the data for all available observed light maxima should display a trend described by:
\begin{displaymath}
\langle u(x) \rangle^2 = 2 a^2 + x e^2 \;,
\end{displaymath}
where {\it a} is the average uncertainty in days for established times of light maxima and {\it e} is the magnitude of any random fluctuations in period. The technique could alternatively be formulated to represent {\it e} in terms of phase offset. A schematic representation of expectations for a ``typical'' pulsating variable is presented in Fig.~\ref{fig2}. But such expectations are never matched exactly in practice, since the dominant pulsation in such stars forces the random factor in the pulsations back into a regular pattern of variability after $\sim 50-200$ cycles.\\[2mm]

\section{Different Types of Pulsating Stars}

  Although the Eddington-Plakidis test was devised specifically to analyze random fluctuations in pulsation period for Mira variables, and later applications by Percy et al. (1993), Percy \& Hale (1998), and Percy \& Colivas (1999) included ``Mira-like'' stars (Miras, Type A and B semiregulars, and RV Turi variables), the same technique should be valid for all types of stars. The technique was also applied to short period pulsators of both Population types by Percy et al. (2003, 2007), was extended to Cepheids in a number of studies (Turner \& Berdnikov 2001, 2004; Abdel-Sabour Abdel-Latif 2004; Berdnikov et al. 2004, 2007, 2009a, 2009b; Berdnikov 2010, Berdnikov \& Stevens 2010). Mostly negative results were found for short period Cepheids, but that is because individual light maxima are rarely observed for such stars and the individual times of light maximum tabulated in O--C analyses usually refer to data obtained over many adjacent cycles about the one cited. Since random fluctuations in period exist over several pulsation cycles, their effects on times of light maximum can easily be confused with other sources of scatter in the light curves when the data are averaged over many cycles.

  Confirmation of that conclusion has recently come from space observations of Cepheids (Berdnikov 2010; Berdnikov \& Stevens 2010). Since weather problems and limited observing windows are generally not a problem for space observations, it is possible to observe many consecutive light maxima for short period Cepheids and to derive observed times of light maxima from applications of Hertzsprung's method. The results indicate that random fluctuations in period are relatively common even for short period Cepheids, and most likely apply to all radially pulsating stars, at least to some extent.

  The same feature also appears in at least one SRC variable, the pulsating M3 Ia supergiant BC Cyg, a star for which the observed times of light maximum can vary by $\pm84$ days from those predicted by a quadratic ephemeris accounting for its long-term period decrease (Turner et al. 2009). An Eddington-Plakidis test for the star is replicated in Fig.~\ref{fig3}, which demonstrates that random fluctuations in pulsation period for the star dominate over small cycle counts, whereas the dominant regular pulsation appears to return the variability to a regular pattern after about 12 cycles or so ($\sim 23$ years in the present case). In pulsating stars of shorter cycle length such regularity returns much sooner, a matter of a few years in the case of short period Cepheids (Berdnikov 2010).

Evidence for random fluctuations in pulsation period on short time scales can also be seen in the results of Poleski (2008) for several Cepheids in the Large Magellanic Cloud. The evidence is revealed by offsets in the observed times of light maximum from cycle to cycle in Poleski's O--C diagrams, although the deviations are typically rather small in comparison with much larger deviations observed over longer time intervals, where the evolutionary changes in mean radius become dominant (Turner et al. 2006).

\section{Parameterizing the Randomness Factor}

  In their original tests on $o$ Ceti (Mira) and $\chi$ Cygni, Eddington \& Plakidis (1929) noted that the observed random fluctuations in period for both stars amounted to about the same amount, 1.35\% to 1.39\% of the pulsation period. They clearly understood the importance of the star's pulsation period to the magnitude of the stochastic processes producing the random fluctuations in period. Yet the cycle length for any pulsating star also depends directly on stellar radius through the period-radius relations applying to every type of pulsating star. Therefore, a better parameter for describing the stochastic processes arising in pulsating stars is the ratio $e/P$, although that is only to first order. The parameter $e/P$ must be independent of radius if the variables obey similar period-radius relations.

\begin{figure}[!t]
\centerline{
\epsfig{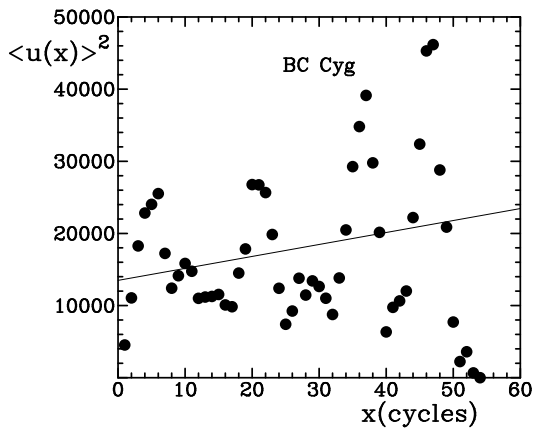}
}
\caption{\small{An Eddington-Plakidis test for the M3 Ia supergiant variable BC Cyg.}}
\label{fig3}
\end{figure}

\begin{figure}[!t]
\centerline{
\epsfig{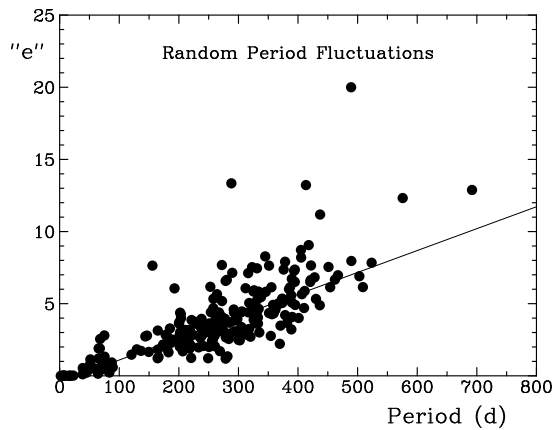}
}
\caption{\small{The observed trend of increasing randomness factor ``{\it e}'' with increasing pulsation period {\it P}.}}
\label{fig4}
\end{figure}

\begin{figure}[!t]
\centerline{
\epsfig{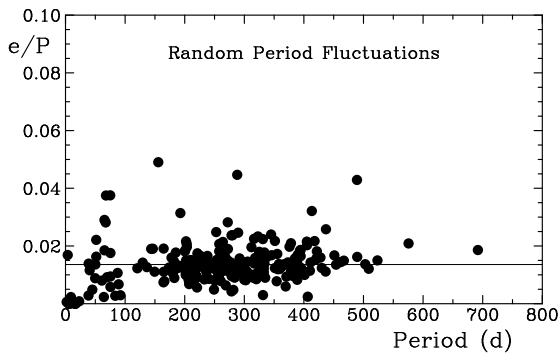}
}
\caption{\small{The nearly negligible trend of the parameter $e/P$ with pulsation period.}}
\label{fig5}
\end{figure}

  Results to date for all Eddington-Plakidis analyses of pulsating stars (updated from Turner et al. 2009) are illustrated in Figs.~\ref{fig4}~and~\ref{fig5}, and confirm such an assumption. Note that Fig.~\ref{fig4} contains additional data from earlier versions plotted by Turner \& Berdnikov (2001) and Turner et al. (2009). As seen in Fig.~\ref{fig5}, the parameter $e/P$ is indeed relatively independent of pulsation period, in other words independent of stellar radius, although there may be an additional trend with period, perhaps reflecting the increasing dominance of envelope convection with decreasing stellar surface temperature in such stars. The parameter $e/P$ has a mean value of $0.0136 \pm0.0005$ ($\pm0.0069$ s.d.), which matches the results of Eddington \& Plakidis (1929) more than 80 years ago.

  The actual variation of randomness factor {\it e} with pulsation period {\it P} may be somewhat more complicated. For example, most of the stars with the smallest values of $e/P$ are short period Cepheids of relatively high mean effective temperature, whereas the star of longest pulsation period, BC Cyg, may have a larger value of $e/P$ than indicated in Figs.~\ref{fig4}~and~\ref{fig5}, if one considers the variation of $\langle u(x) \rangle^2$ over small cycle differences {\it x} in Fig.~\ref{fig3}. The M3 Ia supergiant BC Cyg is, of course, the star of lowest mean effective temperature in the sample.

  As noted earlier, the actual observed trends in the computed values of $\langle u(x) \rangle^2$ for all pulsating stars tested to date initially increase directly in proportion to increasing differences in cycle count {\it x}, as predicted (Eddington \& Plakidis 1929, see also Fig.~\ref{fig2}). At larger cycle differences, however, the trend reverses as the dominant pulsation in such stars reimposes its regularity in the observed times of light maximum. Observed values of $\langle u(x) \rangle^2$ gradually become much smaller for large cycle differences, a characteristic also noted by Eddington and Plakidis in their original study of two Mira variables and seen in almost all of the Eddington-Plakidis tests cited earlier. The stochastic fluctuations in period that appear as a common feature in the cycle lengths of nearly all pulsating stars are therefore dominated by the regular pulsation in such stars. As noted above, the physical processes responsible for such characteristics are uncertain, but presumably originate in temporal modifications of envelope convection in such stars.

\section{Discussion}

  The fact that random fluctuations in period are ubiquitous for all pulsating variables has important consequences. Standard O--C analyses of Cepheids, for example, will always display scatter in the individual O--C data, unless they are averaged over many adjacent cycles. Even then, the non-photometric source of scatter in light curve data points for individual cycles must also introduce a small source of uncertainty in the resulting O--C data based on light curves constructed from observations averaged over the same cycles. Likewise, the detection of sizable random fluctuations in period for long period variables means that the {\it Predicted Dates of Maxima and Minima of Long Period Variables} issued regularly by the AAVSO must necessarily be inexact. Fortunately such predictions are generally issued within a cycle or two of the predicted dates, so they are likely to be only a few days off because of the stochastic processes occurring in the envelopes of such stars.

\begin{figure}[!t]
\centerline{
\epsfig{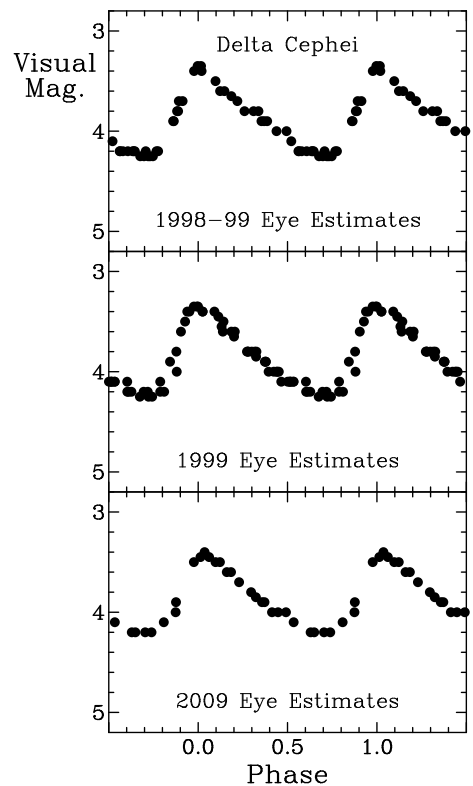}
}
\caption{\small{Observations by Turner with unaided eye of $\delta$ Cephei from 1998 to 2009.}}
\label{fig6}
\end{figure}

  The AAVSO collection of data for Miras and long period variables was the basis for studies of random fluctuations in their pulsation periods by Percy et al. (1993), Percy \& Hale (1998), and Percy \& Colivas (1999), so it seems clear that such changes can be detected in long period variables from simple eye estimates. An interesting question to ask is whether or not such effects can also be detected from simple eye estimates for Cepheid variables. The lead author recently had an opportunity to address that question while teaching an undergraduate course in astronomy. A decade previously, Turner (1999, 2000) demonstrated a simple procedure for obtaining precise estimates of magnitude for bright Cepheids from unaided observations by eye. The original observations from 1998-99 have been used frequently since then for instructional purposes, and the procedure was revived in October 2009 to provide a reference data set for a student attempting to follow the technique on his own. The results are illustrated in Fig.~\ref{fig6}, where the data are phased using the same ephemeris adopted in 1998-99.

  The observed light maxima for $\delta$ Cephei in winter 1998-99 and fall 1999 fell very close to the times predicted from the adopted ephemeris for the star, and the same conclusion applies to the observations from 2009. The Cepheid has a well-established period decrease (Turner 1999, 2000; Turner et al. 2006), so the times of light maximum should occur slightly earlier in 2009 than they did in 1998-99. But the O--C offset is effectively nil between the two dates, and no evidence for chaotic effects in pulsation period can be distinguished from our eye estimates for the star. The results of Berdnikov (2010) and Berdnikov \& Stevens (2010) for other short-period Cepheids are consistent with such conclusions.

\subsection*{\sc{acknowledgements}}
\small{The authors acknowledge with thanks the variable star observations from the American Association of Variable Star Observers (AAVSO) International Database contributed by observers worldwide that are displayed in Fig.~\ref{fig1} of this study.}

\end{document}